\documentclass[graybox]{svmult}

\usepackage{mathptmx}       
\usepackage{helvet}         
\usepackage{courier}        
\usepackage{type1cm}        
\usepackage{makeidx}         
\usepackage{graphicx}        
\usepackage{multicol}        
\usepackage[bottom]{footmisc}

\usepackage{verbatim}
\usepackage{natbib}
\usepackage[utf8]{inputenc}

\makeindex             

\begin{document}

\title*{Why have asset price properties changed so little in 200 years}

\author{Jean-Philippe Bouchaud and Damien Challet}

\institute{Jean-Philippe Bouchaud \at Capital Fund Management, Rue de l'Université 23, 75007 Paris, France; \& Ecole Polytechnique, Palaiseau, France, \email{jean-philippe.bouchaud@cfm.fr}
\and Damien Challet \at Laboratoire de Mathématiques et Informatique pour la Complexité et les Systèmes, CentraleSupélec, Paris, France;  \& Encelade Capital SA, Lausanne, Switzerland, \email{damien.challet@centralesupelec.fr}}

\maketitle

\abstract*{We first review empirical evidence that asset prices have had episodes of large fluctuations and been inefficient for at least 200 years. We briefly review recent theoretical results as well as the neurological basis of trend following and finally argue that these asset price properties can be attributed to two fundamental mechanisms that have not changed for many centuries: an innate preference for trend following and the collective tendency to exploit as much as possible detectable price arbitrage, which leads to destabilizing feedback loops.
}

\abstract{We first review empirical evidence that asset prices have had episodes of large fluctuations and been inefficient for at least 200 years. We briefly review recent theoretical results as well as the neurological basis of trend following and finally argue that these asset price properties can be attributed to two fundamental mechanisms that have not changed for many centuries: an innate preference for trend following and the collective tendency to exploit as much as possible detectable price arbitrage, which leads to destabilizing feedback loops.}

\section{Introduction}
\label{sec:1}

According to mainstream economics, financial markets should be both efficient and stable. Efficiency means that the current asset price is an unbiased estimator of its fundamental value (aka ``right'', ``fair'' or ``true'') price. As a consequence, no trading strategy may yield statistically abnormal profits based on public information. Stability implies that all price jumps can only be due to external news.

Real-world price returns have surprisingly regular properties, in particular fat-tailed price returns and lasting high- and low- volatility periods. The question is therefore how to conciliate these statistical properties, both non-trivial and universally observed across markets and centuries, with the efficient market hypothesis.

The alternative hypothesis is that financial markets are intrinsically and chronically unstable. Accordingly, the interactions between traders and prices inevitably lead to price biases, speculative bubbles and instabilities that originate from feed-back loops. This would go a long way in explaining market crises, both fast (liquidity crises, flash crashes) and slow (bubbles and trust crises). This would also explain why crashes did not wait for the advent of modern HFT to occur:  whereas the May 6 2010 flash crash is well known, the one of May 28 1962, of comparable intensity but with only human traders, is much less known.  

The debate about the real nature of financial market is of fundamental importance. As recalled above, efficient markets provide prices that are unbiased, informative estimators of the value of assets. The efficient market hypothesis is not only intellectually enticing, but also very reassuring for individual investors, who can buy stock shares without risking being outsmarted by more savvy investors.

This contribution starts by reviewing 200 years of stylized facts and price predictability. Then, gathering evidence from Experimental Psychology, Neuroscience and agent-based modelling, it outlines a coherent picture of the basic and persistent mechanisms at play in financial markets, which are at the root of destabilizing feedback loops.

\section{Market anomalies}

Among the many asset price anomalies documented in the economic literature since the 1980s (\cite{Schwert}), two of them stand out:
\begin{enumerate}
\item The Momentum Puzzle: price returns are persistent, i.e., past positive (negative) returns predict future positive (negative) returns.
\item The Excess Volatility Puzzle: asset price volatility is much larger than  that of fundamental quantities 
\end{enumerate}
These two effects are not compatible with the efficient market hypothesis and suggest that financial market dynamics 
is influenced by other factors than fundamental quantities. Other puzzles, such as the ``low-volatility'' and ``quality'' anomalies, are also very 
striking, but we will not discuss them here -- see \cite{ang2009high,baker2011benchmarks,ciliberti2015deconstructing,bouchaud2016qualitystocks} for recent reviews.

\subsection{Trends and bubbles}

\begin{figure}
\begin{center}
\includegraphics[width=0.5\textwidth]{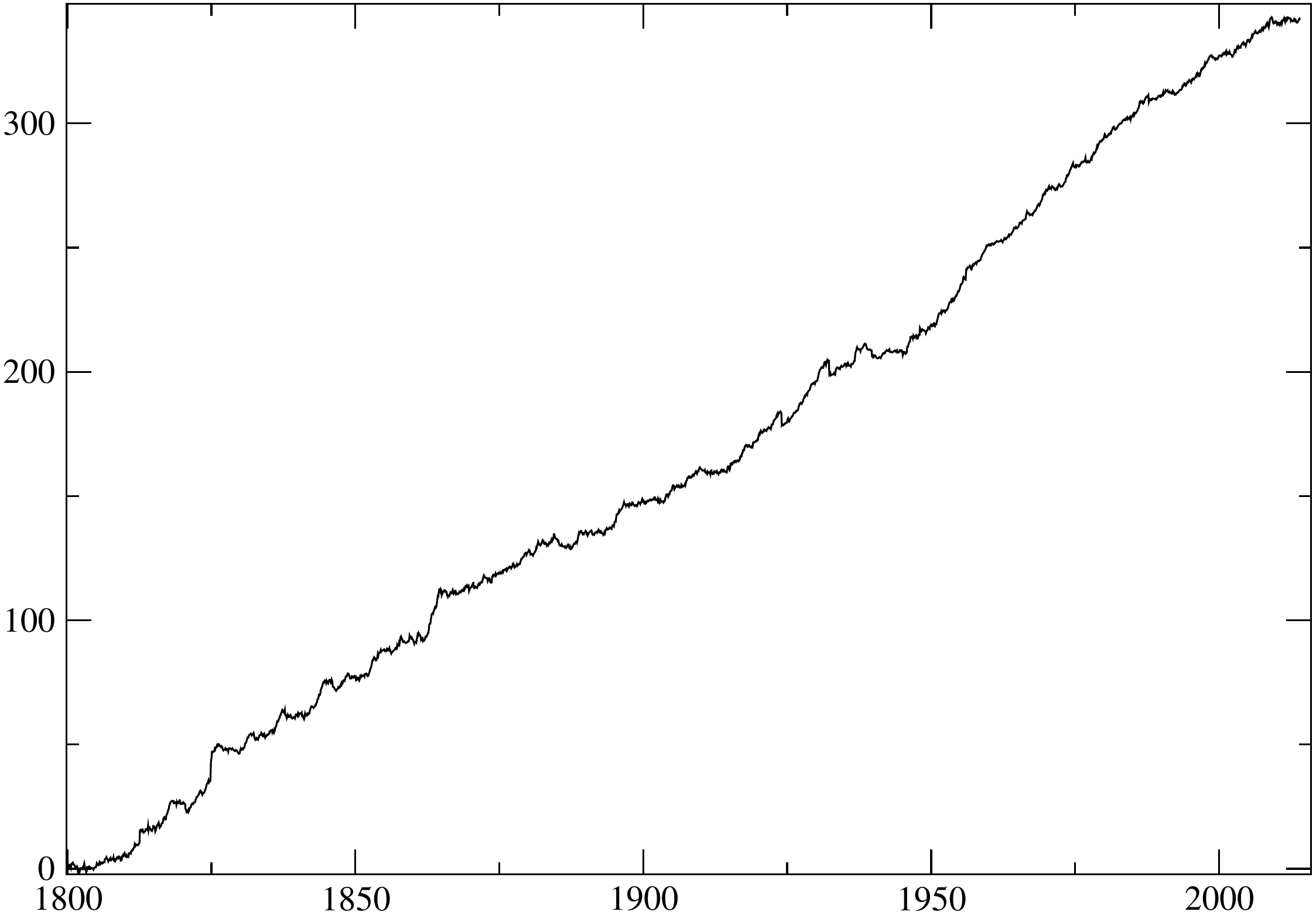}
\caption{Aggregate performance of all sectors of a trend-following strategy with the trend computed over the last six-month moving window, from year 1800 to 2013. 
T-statistics of excess returns is  $9.8$. From \cite{CFM-trends}. Note that the performance in the last 2 years since that study (2014-2015) has been strongly positive.  
}
\label{pnlALLlong}
\end{center}
\end{figure}

In blatant contradiction with the efficient market hypothesis, trend-following strategies have been successful on all asset classes for a very long time. Figure \ref{pnlALLlong} shows for example a backtest of such strategy since 1800 (\cite{CFM-trends}). The regularity of its returns over 200 years implies the presence of a permanent mechanism that makes price returns  persistent. 

Indeed, the propensity to follow past trends is a universal effect, which most likely originates from a behavioural bias: when faced with an uncertain outcome, one is tempted to reuse a simple strategy that seemed to be successful in the past (\cite{Gigerenzer}). The relevance of behavioural biases to financial dynamics, discussed by many authors, among whom Kahneman and Shiller, has been confirmed in many experiments on artificial markets (\cite{smith1988bubbles}), surverys
 (\cite{Shiller_survey, Menkhoff, greenwood2013expectations}), etc. which we summarize in Section \ref{sec:behaviour}.

\begin{figure}
\begin{center}
\includegraphics[width=0.75\textwidth]{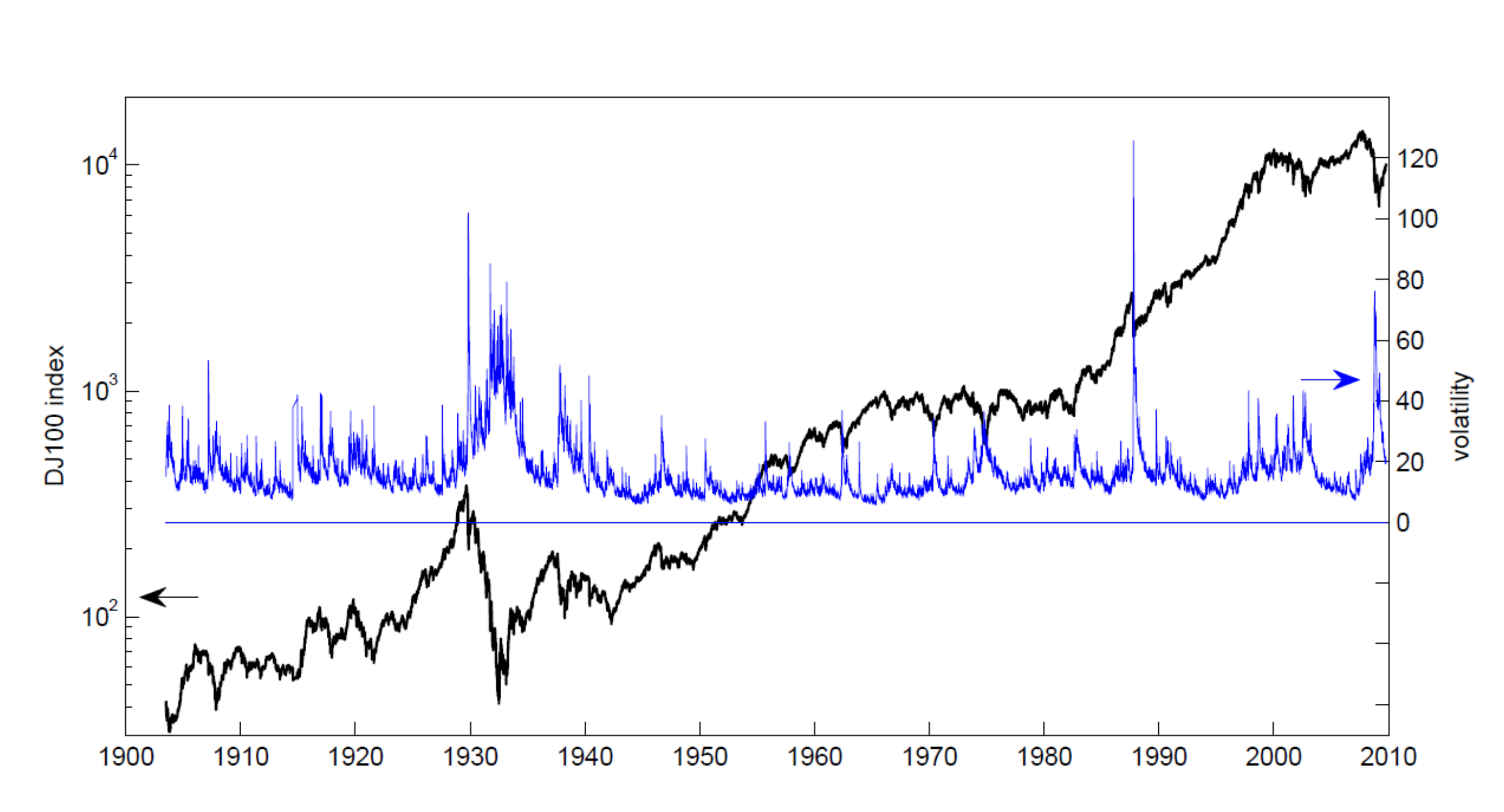}
\caption{Evolution of the Dow-Jones Industrial Average index and its volatility over a century. See \cite{Zumbach}.}
\label{Fig_vol}
\end{center}
\end{figure}

\begin{figure}
\begin{center}
\includegraphics[width=0.75\textwidth]{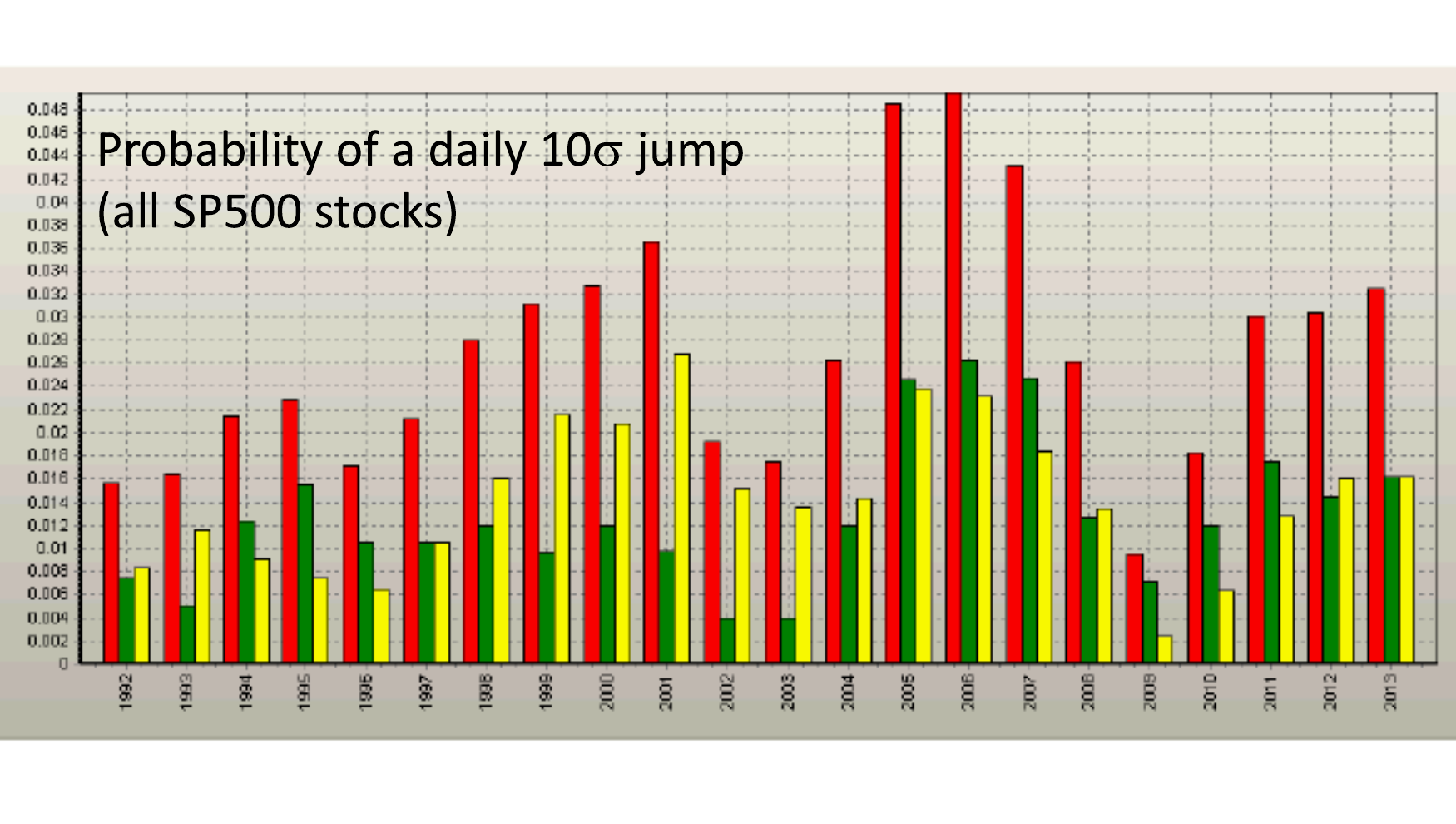}
\caption{Yearly evolution of the probability of the occurence of 10-$\sigma$ price jump for a given day for assets in the S\&P500 since 1991. Yellow bars: positive jumps; green bars: negative jumps; red bars: both of them. These probabilities do vary statistically from year to year, but far less than the volatility itself. This suggests that probability distributions of returns, normalized by their volatility, is universal, even in the tails (cf.\ also Fig.\ \ref{Fig_vol}). Note that the jumps probability has not increased since 1991, despite the emergence of High Frequency Trading (source: Stefano Ciliberti).}
\label{Fig_jumps}
\end{center}
\end{figure}

\subsection{Short-term price dynamics: jumps and endogenous dynamics} 

\subsubsection{Jump statistics}

\begin{figure}
\begin{center}
\includegraphics[width=0.75\textwidth]{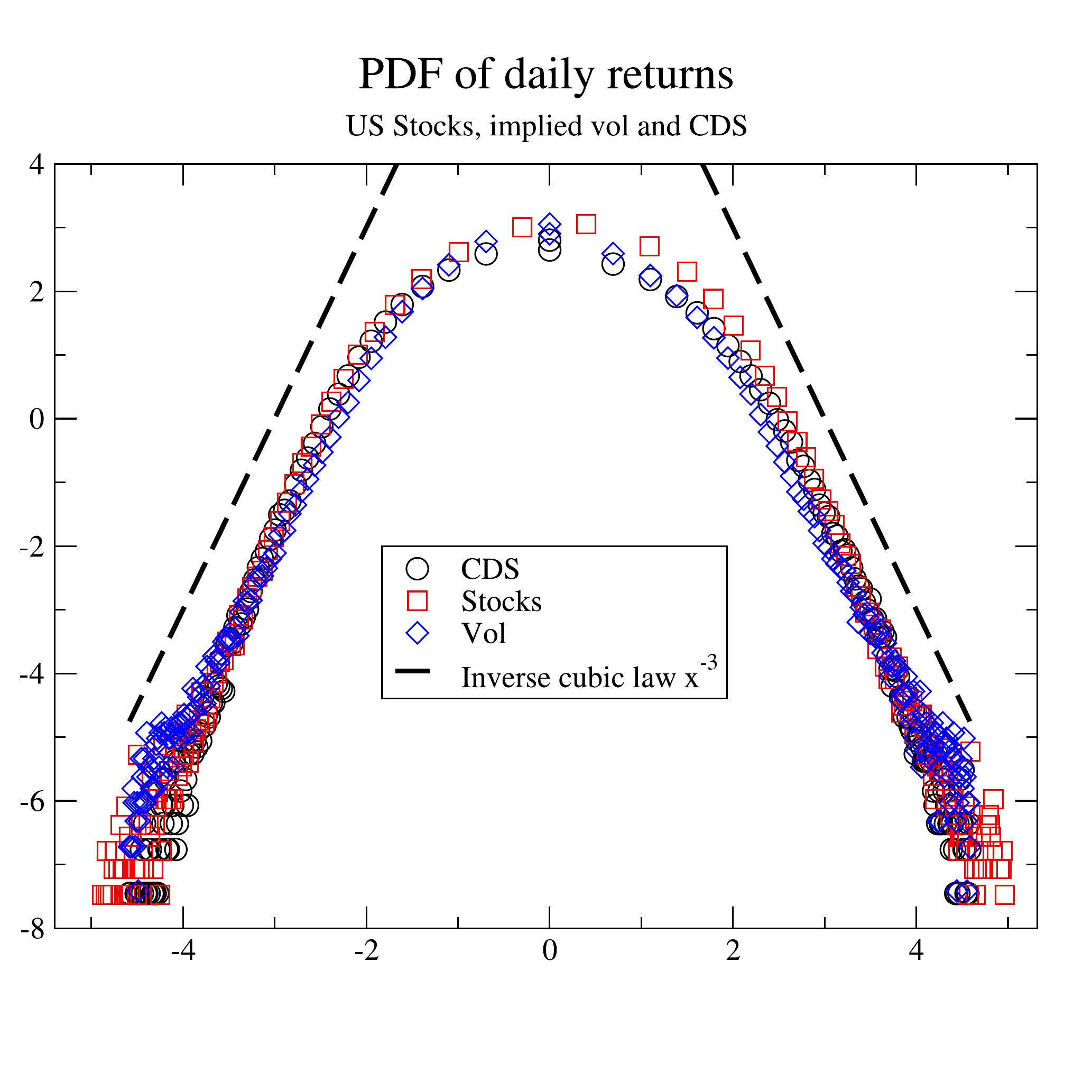}
\caption{Daily price return distributions of price, at-the-money volatility and CDS of the 283 S\& P 500 that have one, between 2010 and 2013.
Once one normalizes the returns of each asset class by their respective volatility, these three distributions are quite similar, despite the fact the asset classes are very different.
The dashed lines correspond to the ``inverse cubic law'' $P(x) \sim |x|^{-1-3}$ (source: Julius Bonart).
\label{Fig_tails}}

\end{center}
\end{figure}

Figure \ref{Fig_tails} shows the empirical price return distributions of assets from three totally different assets classes. The distributions are remarkably similar  (see also \cite{zumbach2015cross}): the probability of extreme return  are all $P(x) \sim |x|^{-1-\mu}$, where the exponent $\mu$ is close to 3 (\cite{Gabaix}). 
The same law holds for other markets (raw materials, currencies, interest rates). This implies that crises of all sizes occur and result into both positive and negative jumps, from fairly small crises to centennial crises.

In addition, and quite remarkably, the probability of the occurence of price jumps is much more stable than volatility (see also \cite{Zumbach}). Figure \ref{Fig_jumps} illustrates this stability by plotting the 10-$\sigma$ price jump probability as a function of time.

\subsubsection{The endogenous nature of price jumps}

What causes these jumps? Far from being rare events, they are part of the daily routine of markets: every day, at least one 5-$\sigma$ event occurs for one of the S\&P500 components! According the Efficient Market Hypothesis, only some very significant pieces of information may cause large jumps, i.e., may substantially change the fundamental value of a given asset. This logical connection is disproved by empirical studies which match news sources with price returns: only a small fraction of jumps can be related to news and thus defined as an exogenous shock (\cite{Poterba,Fair,joulin2008stock,Cornell}).

The inevitable conclusion is that most price jumps are self-inflicted, i.e., are endogenous. From a dynamical point of view, this means that feedback loops are so important that, at times, the state of market dynamics is near critical: small perturbations may cause very large price changes. Many different modelling frameworks yield essentially the same conclusion  (\cite{Wyart,marsili2009dynamic,bacry2012nonparametricHawkes,hardiman2013critical,Remy}). 

The relative importance of exogenous and endogenous shocks is then linked to the propensity of the financial markets to hover near critical or unstable points. The next step is therefore to find mechanisms that systematically tend to bring financial markets on the brink.

\section{Fundamental market mechanisms: arbitrage, behavioural biases and feedback loops}
\label{sec:behaviour}
In short, we argue below that greed and learning are two sufficient ingredients to explain the above stylized facts. There is no doubt that human traders have always tried to outsmart each other, and that the members the {\em homo sapiens sapiens} clique have some learning abilities. Computers and High Frequency Finance then merely decrease the minimum reaction speed (\cite{hardiman2013critical}) without modifying much the essence of the mechanisms at play.

In order to properly understand the nature of the interaction between investors in financial markets, one needs to keep two essential ingredients

\begin{enumerate}

\item Investor heterogeneity: the distribution of their wealth, trading frequency, computing power, etc. have heavy tails, which prevents a representative agent approach. 
\item Asynchronism: the number of trades per agent in a given period is heavy-tailed, which implies that they do not trade synchronously. In addition, the continuous double auction mechanism implies sequential trading: only two orders may interact at any time.

\end{enumerate}

One thus cannot assume that all the investors behave in the same way, nor that they can be split into two or three categories, which is nevertheless a common assumption when modelling or analyzing market behaviour.

\subsection{Speculation}

Although the majority of trades are of algorithmic nature nowadays, most traders (human or artificial) use the same types of strategies. Algorithmic trading very often simply implements analysis and extrapolation rules that have been used by human traders since immemorial times, as they are deeply ingrained in human brains.

\subsubsection{Trend following}

Trend-following in essence consists in assuming that future price changes will be of the same sign as last past price changes.
It is well-known that this type of strategy may destabilize prices by increasing the amplitude and duration of price excursions. Bubbles also last  longer because of heavy-tailed trader heterogeneity. Neglecting new investors for the time being, the heavy-tailed nature of trader reaction times implies that some traders are much slower than others to take part to a nascent bubble. This causes a lasting positive volume imbalance that feeds a bubble for a long time. Finally, a bubble attracts new investors that may be under the impression that this bubble grow further. The neuronal processes that contribute the emergence and duration of bubbles are discussed in section \ref{s-s-section:neurofinance}.

\subsubsection{Contrarian behaviour}

Contrarian trading consists in betting on mean-reverting behavior: price excursions are deemed to be only temporary, i.e., that the price will return to some reference (``fundamental'' or other) value. Given the heterogeneity of traders, one may assume that the do not all have the same reference value in mind.
The dynamical effects of this type of strategies is to stabilize price (with respect to its perceived reference value). 

\subsubsection{Mixing trend followers and contrarians}

In many simplified agent-based models (\cite{degrauwe1993exchange,brock1998heterogeneous,LuxMarchesi99}) both types of strategies are used by some fractions of the trader populations. A given trader may either always use the same kind of strategy (\cite{frankel1986understanding,frankel1990chartists}), may switch depending on some other process (\cite{kirman1991epidemics}) 
or on the recent trading performance of the strategies (\cite{brock1998heterogeneous,wyart2007self,LuxMarchesi99}).  In a real market, the relative importance of a given type of strategy is not constant, which influences the price dynamics. 

Which type of trading strategy dominates can be measured in principle. Let us denote the price volatility measured over a single time step by $\sigma_1$. If trend following dominates, the volatility of returns measured every $T$  units of time, denoted by $\sigma_T$ will be larger than $\sigma_1\sqrt{T}$. Conversely, if mean-reverting dominates, $\sigma_T<\sigma_1\sqrt{T}$. Variance-ratio tests, based on the quantity $\sigma_T/(\sigma_1\sqrt{T})$, are suitable tools to assess the state of the market (see \cite{charles2009varianceratio} for a review); see for example the PUCK concept, proposed by \cite{mizuno2007analysis}.

When trend following dominates, trends and bubbles may last for a long time. The bursting of a bubble may be seen as mean-reversion taking (belatedly)  over. This view is too simplistic, however, as it implicitly assumes that all the traders have the same calibration length and the same strategy parameters. In reality, the periods of calibration used by traders to extrapolate price trends are very heterogeneous. Thus, strategy heterogeneity and the fact that traders have to close their positions some time imply that a more complex analysis is needed.

\subsection{Empirical studies}
In order to study the behaviour of individual investors, the financial literature makes use of several types of data

\begin{enumerate}
\item Surveys about individual strategies and anticipation of the market return over the coming year (\cite{Shiller_survey,greenwood2013expectations}).
\item The daily investment flows in US securities of the sub-population of individual traders. 
The transactions of individual traders are labelled as such, without any information about the identity of the investor (\cite{kaniel2008individual}).
\item  The daily net investment fluxes of each investor in a given market. For example, \cite{tumminello2011identification} 
use data about Nokia in the Finish stock exchange.
\item Transactions of all individual investors of a given broker (\cite{dorn2008correlated,morton2010turnover}). The representativity of such kind of data may be however uestionned (cf.\ next item)
\item Transactions of all individual investors of all the brokers accessing a given market. \cite{jackson2004aggregate} shows that the behaviour of individual investors is the same provided that they use an on-line broker.
\end{enumerate}

\subsubsection{Trend follower vs contrarian}

Many surveys show that institutional and individual investors expectation about future market returns are trend-following (e.g. \cite{greenwood2013expectations}), yet the analysis of the individual investors' trading flow at a given frequency (i.e. daily, weekly, monthly) invariably point out that their actual trading is dominantly contrarian as it is anti-correlated with previous price returns, while institutional trade flow is mostly uncorrelated with recent price changes on average (\cite{grinblatt2000investment,jackson2004aggregate,dorn2008correlated,lillo2008specialization,challet2013robust}), . In addition, the style of trading of a given investor only rarely changes  (\cite{lillo2008specialization}).

Both findings are not as incompatible as it seems, because the latter behaviour is consistent with price discount seeking. In this context, the contrarian nature of investment flows means that individual investors prefer to buy shares of an asset after a negative price return and to sell it after a positive price return, just to get a better price for their deal. If they neglect their own impact, i.e., if the current price is a good approximation of the realized transaction price, this makes sense. If their impact is not negligible, then the traders buy when their expected transaction price is smaller than the current price and conversely (\cite{batista2015investors}).

\subsubsection{Herding behaviour}

\cite{lakonishok1992impact}  define a statistical test of global herding. US mutual funds do not herd, while individual investors significantly do (\cite{dorn2008correlated}).
Instead of defining global herding, \cite{tumminello2011identification} define sub-groups of invidivual investors defined by the synchronization of their activity and inactivity, the rationale being that people that use the same way to analyse information are likely to act in the same fashion. This in fact defines herding at a much more microscopic level. The persistent presence of many sub-groups sheds a new light on herding. Using this method, \cite{challet2016leadlag} show that synchronous sub-groups of institutional investors also exist.

\subsubsection{Behavioural biases}

Many behavioural biases have been reported in the literature. Whereas they are only relevant to human investors, i.e. to individual investors, most institutional funds are not (yet) fully automated and resort to human decisions. We will mention two of the most relevant biases.

Human beings react different to gains and to losses (see e.g. Prospect Theory \cite{KahnemanProspect}) 
and prefer positively skewed returns to negatively skewed returns (aka the ``lottery ticket'' effect, see \cite{lemperiere2016risk}). This has been linked to the disposition bias, which causes investors to close too early winning trades and too late losing ones (\cite{shefrin1985disposition,odean1998investors,boolell2009disposition}) (see however  \cite{ranguelova2001disposition,barberis2009drivesdisposition,annaert2008disposition}). 
An indisputable bias is overconfidence, which leads to an excess of trading activity, which diminishes the net performance \cite{barber2000trading}, see also \cite{batista2015investors} for a recent experiment eliciting this effect. This explains why male traders earn less than female trades \cite{barber2001boys}. Excess confidence is also found in individual portfolios, which are not sufficiently diversified. For example, individual traders trust too much their asset selection abilities (\cite{goetzmann2005individual,calvet2007down}).

\subsection{Learning and market instabilities}

Financial markets force investors to be adaptive, even if they are not always aware of it  (\cite{FarmerForce,ZMEM,LoAMH}).
Indeed, strategy selection operates in two distinct ways
\begin{enumerate}
\item  Implicit: assume that an investor always uses the same strategy and never recalibrates its parameters. The performance of this strategy modulates the wealth of the investor, hence its relative importance on markets. In the worst case, this investor and his strategy effectively disappears. This is the argument attributed to Milton Friedman according to which only rational investors are able to survive in the long run because the uninformed investors are weeded out.
\item Explicit: investors possess several strategies and use them in an adaptive way, according to their recent success. In this case, strategies might die (i.e., not being used), but investors may survive.
\end{enumerate}

The neo-classical theory assumes the convergence of financial asset prices towards an equilibrium in which prices are no longer predictable. The rationale is that market participants are learning optimally such that this outcome is inevitable. A major problem with this approach is that learning requires a strong enough signal-to-noise ratio (Sharpe ratio); as the signal fades away, so does the efficiency of any learning scheme. As a consequence, reaching a perfectly efficient market state is impossible in finite time.

This a major cause of market instability. \cite{patzelt2011criticality} showed that optimal signal removal in presence of noise tends to converge to a critical state characterized by explosive and intermittent fluctuations, which precisely correspond to the stylized facts described in the first part of this paper. This is a completely generic result and directly applies to financial markets. Signal-to-noise mediated transitions to explosive volatility is found in agent-based models in which predictability is measurable, as in the Minority Game  (\cite{CM03,MGbook}) and more sophisticated models (\cite{BouchaudGiardinaCrash}).

\subsection{Experiments}

\subsubsection{Artificial assets}

In their famous work, \cite{smith1988bubbles} found that price bubbles emerged in most experimental sessions, even if only three or four agents were involved. This means that financial bubble do not need very many investors to appear. Interestingly, the more experienced the subjects, the less likely the emergence of a bubble.

More recently, \cite{hommes2005coordination} observed that in such experiments, the resulting price converges towards the rational price either very rapidly or very slowly or else with large oscillations. \cite{anufriev2009evolutionary} assume that the subjects dynamically use very simple linear price extrapolation rules (among which trend-following and mean-reverting rules),
 
\subsubsection{Neurofinance}
\label{s-s-section:neurofinance}

Neurofinance aims at studying the neuronal process involved in investment decisions (see \cite{lo2011fear} for an excellent review). One of the most salient result is that, expectedly, human beings spontaneously prefer to follow perceived past trends.

Various hormones play a central role in the dynamics of risk perception and reward seeking, which are major sources of positive and negative feedback loops in Finance. Even better, hormone secretion by the body modifies the strength of feedback loops dynamically, and feedback loops interact between themselves. Some hormones have a feel-good  effect, while other reinforce to risk aversion.

\cite{coates2008endogenous} measured the cortisol (the ``stress hormone'') concentration in saliva samples of real traders and found that it depends on the realized volatility of their portfolio. This means that a high volatility period durable increases the cortisol level of traders, which increases risk aversion and reduces activity and liquidity of markets, to the detriment of markets as a whole.

Reward-seeking of male traders is regulated by testosterone. The first winning round-trip leads to an increase of the level testosterone, which triggers the production of dopamine, a hormone related to reward-seeking, i.e. of another positive round-trip in this context. This motivates the trader to repeat or increase his pleasure by taking additional risk. At relatively small doses, this exposure to reward and reward-seeking has a positive effect. However, quite clearly, it corresponds to a destabilizing feedback loop and certainly reinforces speculative bubbles.
Accordingly, the trading performance of investors is linked to their dopamine level, which is partly determined by genes (\cite{lo2005fear,sapra2012combination}).
.

Quite remarkably, the way various brain areas are activated during the successive phases of speculative bubbles has been investigated in detail.
\cite{lohrenz2007fictive} suggest a neurological mechanism which motivates investors to try to ride a bubble: they correlate the activity of a brain area with how much gain opportunities a trader has missed since the start of a bubble. This triggers the production of dopamine, which in turn triggers risk taking, and therefore generates trades. In other words, regrets or ``fear of missing out'' lead to trend following. 

After a while, dopamine, i.e., gut feelings, cannot sustain bubbles anymore as its effect fades. Another cerebral region takes over;  quite ironically, it is one of the more rational ones:
\cite{demartino2013theorymind} find a correlation between the activation level of an area known to compute a representation of the mental state of other people, and the propensity to invest in a pre-existing bubble. These authors conclude that investors make up a rational explanation about the existence of the bubble (``others cannot be wrong'') which justifies to further invest in the bubble. This is yet another neurological explanation of our human propensity to trend following.

\subsection{Conclusion}

Many theoretical arguments suggest that volatility bursts may be intimately related to the quasi-efficiency of financial markets, in the sense that predicting them is hard because the signal-to-noise ratio is very small (which does not imply that the prices are close to their ``fundamental'' values). Since the adaptive behaviour of investors tends to remove price predictability, which is the signal that traders try to learn, price dynamics becomes unstable as they then base their trading decision on noise only (\cite{MGbook,patzelt2011criticality}). This is a purely endogenous phenomenon whose origin is the implicit or explicit learning of the value of trading strategies, i.e., of the interaction between the strategies that investors use. This explains why these stylized facts have existed for at least as long as financial historical data exists. Before computers, traders used their strategies in the best way they could. Granted, they certainly could exploit less of the  signal-to-noise ratio than we can today. This however does not matter at all: efficiency is only defined with respect to the set of strategies one has in one's bag. As time went on, the computational power increased tremendously, with the same result: unstable prices and bursts of volatility. This is why, unless exchange rules are dramatically changed, there is no reason to expect financial markets will behave any differently in the future. 

Similarly, the way human beings learn also explains why speculative bubbles do not need rumour spreading on internet and social networks in order to exist.  
Looking at the chart of an asset price is enough for many investors to reach similar (and hasty) conclusions without the need for peer-to-peer communication devices (phones, emails, etc.). In short, the fear of missing out is a kind of indirect social contagion.

Human brains have most probably changed very little for the last two thousand years. This means that the neurological mechanisms responsible for the propensity to invest in bubbles are likely to influence the behaviour of human investors for as long as they will be allowed to trade. 

From a scientific point of view, the persistence of all the above mechanisms justifies the quest for the fundamental mechanisms of market dynamics. We believe that the above summary provides a coherent picture of how financial markets have worked for at least two centuries (\cite{Rogoff}) and why they will probably continue to stutter in the future.

\bibliographystyle{plainnat}
\bibliography{biblio.bib}
\end{document}